\documentclass[12pt,amsfonts]{article}    
\usepackage{amsfonts}

\def\one{1\hskip-.37em 1}

\def\ir{{\rm I}\hskip-.2em{\rm R}}

\def\vp{\varphi}
\def\bl{\bf{\l}}

\def\half{\textstyle{\frac{1}{2}}}

\def\halfnu{\textstyle{\frac{1}{2\nu}}}
\def\iN{{\rm I}\hskip-.2em{\rm N}}
\def\iH{{\rm I}\hskip-.2em{\rm H}}
\def\H{{\cal H}}
\def\ZZ{\cal Z}
\def\mfH{\mathfrak{H}}
\def\threebytwo{\textstyle{\frac{3}{2}}}
\def\eigth{\textstyle{\frac{1}{8}}}
\def\EE{{\cal E}}
\def\p{\phi}

\def\th{\theta}
\def\H{{\cal H}}
\def\B{\beta}
\def\v{\vskip.3cm}
\def\g{\gamma}
\def\kk{k^{\sf plus}}
\def\l{\lambda}
\def\D{{\cal D}}
\def\si{\sigma}
\def\S{\Sigma'}
\def\t{\textstyle}
\def\te{}  
\def\De{\Delta}
\def\F{{\cal F}}
\def\tr{\rm Tr}
\def\E{{\rm I}\hskip-.2em{\rm E}}
\def\ra{\rightarrow}
\def\tint{{\textstyle\int}}
\def\hg{{\hat g}}
\def\hp{{\hat\pi}}
\def\hph{{\hat\phi}}
\def\s{\hskip.08em}
\def\P{\Pi'}
\def\d{\partial}
\def\o{\overline}
\def\a{\alpha}
\def\b{\begin{eqnarray*}}  
\def\e{\end{eqnarray*}}    
\def\bn{\begin{eqnarray}}  
\def\en{\end{eqnarray}}   

\def\<{\langle}
\def\>{\rangle}
\def\g{\gamma}
\def\bk{\mathbf k}
\def\bm{\mathbf m}
\def\dn{d^n\!x}
\def\no{\nonumber}

\def\ds{d^s\!x}
\def\k{\kappa}
\def\bl{\bold l}
\def\quarter{\textstyle{\frac{1}{4}}}

\def\hk{\hat{\kappa}}
\def\{{\lbrace}
\def\}{\rbrace}
\begin{document}

\title{Quantum Gravity Made Easy}          
\author{John R. Klauder\footnote{klauder@
phys.ufl.edu} \\
Department of Physics and Department of Mathematics \\
University of Florida,   
Gainesville, FL 32611-8440}
\date{ }
\bibliographystyle{unsrt}
\maketitle
\begin{abstract}
Gravity does not naturally fit well with canonical quantization. Affine quantization is an alternative procedure that is similar to canonical quantization but may offer a positive result when canonical quantization fails to offer a positive result. Two basic examples given initially illustrate the power of affine quantization. These examples clearly point toward an affine quantization procedure that vastly simplifies a successful quantization of the most difficult part of quantum general relativity.  \end{abstract}%

\section{Introduction}

In order to offer a credible analysis of quantum gravity it is first necessary
to carefully review several common questions: (1) Are the rules of canonical quantization the full story of how to quantize any particular classical theory? (2) Is the standard assumption that the  correct  set of basic, phase space classical variables to promote to operator variables
are Cartesian coordinates? (3) How do we choose Cartesian, phase space coordinates when phase space has no metric? (4) Is it necessary  when taking the classical limit of a quantum theory to choose   $\hbar\ra0$ while the classical world around us chooses $\hbar>0$?
\subsection{From canonical to affine vaiables}

Canonical quantization generally works well, but it turns out that there is more to the story that can help when problems arise. We start simply.
For a single degree of freedom, canonical quantization involves $Q$ and $P$ which (ideally) are self-adjoint operators that satisfy the commutator $[Q,P]=i\hbar\mathbb{I}$.
It automatically follows that 
 \bn
   Q\,[Q,P]\hskip-1em&& =[Q,QP]=[Q,(QP+PQ)+(QP-PQ)]/2\no \\
   &&=[Q,(QP+PQ)/2]\equiv [Q,D]=i\hbar Q\,. \label{e1}
   \en
  As usual, $Q$ and $P$ are irreducible, but $Q$ and $D$ are reducible in that
  $D$ and $Q>0$ is irreducible along with $D$ and $Q<0$; a third case where $Q=0$ is less important.
If $Q>0$ (or $Q<0$), then $P$ cannot be made self adjoint; however, in that case both $Q$ and $D$ are self adjoint.
The operator $D$ is called the dilation operator because it {\it dilates} $Q$ rather than {\it translates} $Q$ as $P$ does; in particular, 
   \bn  e^{iqP/\hbar}\,Q\,e^{-iqP/\hbar}=Q+q\mathbb{I} \,\,, 
     \;\; e^{i\ln(|q|)D/\hbar}\,Q\,e^{-i\ln(|q|)D/\hbar}=|q|Q =q|Q|\,,\en
     where in the second relation $q\neq0$, and $q$, as well as $Q$, are normally  chosen to be dimensionless.
     
     Observe: {\it According to (\ref{e1}), the existence of canonical operators {\bf guarantees} the existence of affine operators!}
     
     \subsection{Canonical and affine coherent states}
     The canonical coherent states are well known and generally given by
     \bn  |p,q\rangle \equiv e^{-iqP/\hbar}e^{ipQ/\hbar}|0\rangle\,,\en
     where the fiducial vector $|0\>$ satisfies $(\omega Q+iP)|0\rangle=0$. These vectors admit a resolution of unity given by
        \bn \mathbb{I}=\textstyle{\int}|p,q\rangle\langle p,q|\, dp\,dq/2\pi\hbar \,.\en
        
  The affine coherent states are less well known and they are generally given, for $q>0$ and $Q>0$, by
  \bn |p,q\rangle\equiv e^{ipQ/\hbar} e^{-i\ln(q)D/\hbar}|\beta\rangle\,,\en
  where the fiducial vector $|\beta\>$ satisfies $[(Q-1)+(iD/\beta)]|\beta\rangle =0$; we choose a common notation for 
  the two sets of coherent states, but the different range of variables helps set them apart. The affine vectors
  admit a resolution of unity given by
  \bn  \mathbb{I}=\textstyle{\int}|p,q\rangle\langle p,q| \,dp\,dq/2\pi\hbar C\,,\en
  where $C=[1-\hbar/2\beta]^{-1}$, which requires that $\beta>\hbar/2$.
  A similar story applies to $q<0$ and $Q<0$, or a combination so that
 $q\neq0$, but we focus on $q>0$ which has more relevance for gravity.
 
 \subsection{Classical/quantum connection}
 The connection between classical and quantum variables, while sometimes difficult in conventional canonical 
 quantization, has a clear relationship in the program of Enhanced Quantization \cite{eq}.
 Schr\"odinger's equation for canonical quantization arises from stationary variations of the normalized Hilbert state vectors $|\psi(t)\rangle$ in the
 action functional 
    \bn A_Q=\textstyle{\int}_0^T\,\langle\psi(t)|[i\hbar(\partial/\partial t)-
    \mfH(P,Q)\,]|\psi(t)\rangle\;dt\,,\en
    and the variational result is given by
     \bn i\hbar\,\partial |\psi(t)\rangle/\partial t= \mfH(P,Q)|\psi(t)\rangle\;.\en 
     
    A similar study applies to affine quantization for which Schr\"odinger's equation arises by stationary variations of normalized Hilbert space vectors from the action functional
   \bn A'_Q=\textstyle{\int}_0^T\,\langle\psi(t)|[i\hbar(\partial/\partial t)-\mfH'(D,Q]|\psi(t)\rangle\;dt\,,\en
   and the variational result is given by
    \bn i\hbar\,\partial |\psi(t)\rangle/\partial t= \mfH'(D,Q)|\psi(t)\rangle\;.\en
    
    Classical observers, however, can not explore all the variations that lead to Schr\"odinger's equation. In particular, allowed variations involve simple translations and constant velocities which, according to Galileo invariance, can be made by moving the observer rather than moving the object. Using canonical coherent states, the reduced (R) action functional leads to
    \bn A_{Q(R)}\hskip-1em&&=\textstyle{\int}_0^T\langle p(t),q(t)|[i\hbar(\partial/
    \partial t)-\mfH(P,Q)]|p(t),q(t)\rangle \;dt\no \\
       &&=\textstyle{\int}_0^T [p(t)\dot{q}(t)-H(p(t),q(t)]\;dt\;.\label{ca} \en
 For the affine story we use affine coherent states which lead to
     \bn A'_{Q(R)}\hskip-1em&&=\textstyle{\int}_0^T\langle p(t),q(t)|[i\hbar(\partial/
    \partial t)-\mfH'(D,Q)]|p(t),q(t)\rangle \; dt\no\\
       &&=\textstyle{\int}_0^T[-q(t)\dot{p}(t)-H'(p(t),q(t)]\;dt\;.  \label{af} \en
       Equation (\ref{af}) applies as well when $q\neq 0$, which makes it more similar 
       to Eq. (\ref{ca}).
       
       Notice that the canonical and affine versions of the reduced action functionals are effectively {\it identical} in that they both appear as classical action functionals! In fact, they are `better' than classical expressions because they still involve  $\hbar$ which is not zero. To recover the usual classical story from the quantum story in conventional canonical quantization requires that $\hbar\rightarrow0$,
       but that is highly unphysical because the world we all live in is one where $\hbar>0$. Indeed,
       we prefer to refer to the Hamiltonians in (\ref{ca}) and (\ref{af})) as {\it enhanced classical Hamiltonians} because they each retain $\hbar>0$.
       
      The expressions for the enhanced classical actions for both the canonical and affine stories have the property that if phase space coordinates are changed, such as $(p,q)\ra(\bar{p},\bar{q})$, where
      $p\,dq=\bar{p}\,d\bar{q}+d\bar{G}(\bar{p},\bar{q})$,
      the coherent state vectors satisfy  $|\bar{p},\bar{q}\>=|p,q\>$
      because the original point in phase space must be mapped to the same vector in Hilbert space. This property ensures that even though the phase space variables are changed all of the quantum aspects remain unchanged; this 
      favorable property works whether the underlying operators are canonical or affine.
       
     {\it The common behavior of the enhanced classical stories implies that a classical theory can be quantized by either canonical or affine procedures with the same justification. If one approach fails, try the other one!}
     
     \subsection{ ``Cartesian coordinates"}
     Canonical quantization `promotes' classical variables to operators, e.g., $p\rightarrow P$ and $q\rightarrow Q$,
     and builds its operator Hamiltonian from $H(p,q)\rightarrow\mfH(P,Q)$. But which pair of
     classical variables should be promoted to operators. The standard answer to this question is that the proper classical phase space variables should be {\it `Cartesian coordinates'}, according to Dirac \cite{dirac} (page 114, in a footnote). 
     
     Enhanced quantization offers a clear connection of quantum and classical variables. For the canonical case, the enhanced classical Hamiltonian is given by
       \bn H(p,q)\hskip-1em&&=\<p,q|\,\mfH(P,Q)\,|p,q\>\no\\
            &&=\<0|\mfH(P+p,Q+q)|0\>\no\\
            &&=\mfH(p,q)+{\cal{O}}(\hbar;p,q)\;.  \en
 If, for clarity, $\mfH$ is a polynomial, $\hbar\ra0$, and (c) refers to the normal
 classical limit, it follows that $H_c(p,q)=\mfH(p,q)$, i.e.,  the quantum function of variables should follow the classical function of variables, which, initially, points {\it exactly} toward the goal of choosing `Cartesian coordinates'!
     
     Phase space has no metric by which to determine Cartesian coordinates so that may cause problems. However, Hilbert space has a metric which can be used to examine the question. Consider the Fubini-Study metric \cite{FS} for the canonical coherent states that evaluates the distance-squared between
     two infinitesimally close ray-vectors (minimized over any simple phase), which leads to
  \bn d\sigma(p,q)^2\equiv 2\hbar[\,|\!|\,d|p,q\rangle|\!|^2-
  |\langle p,q|\,d| p,q\rangle|^2 ]=\omega^{-1}dp^2+\omega dq^2 \;,\en
         {\it and offers a flat space that already involves Cartesian coordinates, thereby confirming Dirac's rule!}
         
         For the case of affine variables, with $q>0$ as our example, the enhanced classical Hamiltonian is given by
         \bn H(p,q)\equiv H'(pq,q)\hskip-1em&& =\<p,q|\mfH'(D,Q)|p,q\>\no\\
         &&=\<\beta|\mfH'(D+pqQ, qQ)|\beta\> \\
         &&=\mfH'(pq,q)+{\cal{O}}(\hbar;p,q)\;. \no \en
         Thus, with a polynomial Hamiltonian for clarity, and in the normal classical limit (c) when $\hbar\ra0$, it follows that             
                   \bn H'_c(pq,q)=\mfH'(pq,q)\;.\label{afcool} \en
         {\it Equation (\ref{afcool}) establishes the fact that the quantum function of affine quantum variables should be the same as the classical function of classical affine variables, a connection for affine quantization that is the analog of Dirac's initial rule for canonical quantization!}
          
         In the affine, the  Fubini-Study metric leads to 
  \bn d\sigma(p,q)^2\equiv 2\hbar[\,|\!|\,d|p,q\rangle|\!|^2-
  |\langle p,q|\,d| p,q\rangle|^2]=\beta^{-1}q^2\,dp^2+\beta q^{-2}\,dq^2  \;,\label{yes} \label{22} \en 
  {\it which describes a Poincar\'e half plane\footnote{Expressed here in coordinates where $q\rightarrow {\beta}/q$.}, has a constant negative curvature, $-2/\beta$, and is geodesically complete} \cite{p}! Observed that the `quadratic coefficient' terms in (\ref{22}) yield ${\it unity}$ when they are multiplied together; this property for affine metrics will arise again.
     
     The simple expressions of the two metrics is partly due to the choice of fiducial vectors.
     However, more general fiducial vectors still lead to fairly simple metric expressions.
     
     Do affine spaces of constant negative curvature have `favored coordinates' for quantization like the Cartesian coordinates in flat space? A positive answer to that question follows from the fact that the chosen coordinates of the constant negative curvature space in (\ref{yes}) are the same coordinates in which (\ref{afcool}) holds true, and those coordinates also 
     pass directly to the favored Cartesian coordinates in flat space as $q\rightarrow q+(\beta/\omega)^{1/2}$ and $\beta \rightarrow\infty$.\footnote{Incidentally, a similar story of favored coordinates applies to three-dimensional spin coherent states with a metric in which a spherical space with a constant {\it positive} curvature passes to a metric in a flat space expressed in favored Cartesian coordinates as the spin value $s\ra\infty$ \cite{eq}.}

     {\it Note that the phase space metrics  arose from a Hilbert space and not from the phase space.     Nevertheless, the metrics may be added to the phase space if one chooses.}

     \subsection{An example}
     The harmonic oscillator with the classical Hamiltonian $H(p,q)=\half[p^2+q^2]$, where
     $(p,q)\in\mathbb{R}^2$, and its canonical quantization is so well known we rely on the reader for its behavior; moreover, for this example, we concede that canonical quantization beats affine quantization. However, the identical classical Hamiltonian which is now 
     restricted so that $q>0$, i.e., $(p,q)\in\mathbb{R}\times\mathbb{R}^+$, cannot be correctly quantized by canonical procedures because the proposed operator $P$ cannot be made self adjoint, and $P^\dagger$ has a larger domain than $P$. As one set of possibilities, this leads to $\half[p^2+q^2]$ becoming either $\half[P^\dagger P+Q^2]$ or $\half[P P^\dagger+Q^2]$.\footnote{The choice $\half[P^\dagger P^\dagger+Q^2]$ and its adjoint (both with $Q>0$) are not self-adjoint and may have complex eigenvalues. The sum of these two examples, divided by 2, is 
     self-adjoint and even has the correct classical limit;
      we leave its quantization as an exercise for the reader.}
     The first version has eigenfunctions for (the positive half of) {\it odd} Hermite functions and eigenvalues given by $\hbar[(1, 3, 5, 7, \cdots)+1/2]$ while the second version has eigenfunctions for (the positive half of) {\it even} Hermite functions and eigenvalues given $\hbar[((0, 2, 4, 6, \cdots)+1/2]$, either form being acceptable. Moreover, a valid quantization example could have {\it mixed} eigenvalues such as $\hbar[(0, 1, 3, 4, 6, 8, 9, \cdots)+1/2]$, which, for arbitrary mixing, implies there are infinitely many different mixed solutions offered by canonical quantization. Clearly, infinitely many mixed solutions is not acceptable!
     
     Let us try affine quantization. The classical affine variables are $d=pq$ and $q$. Thus the classical Hamiltonian is now given as $H'(d,q)=\half[dq^{-2}d+q^2]$, and its affine quantization is given by
       $\mfH'(D,Q)=\half[DQ^{-2}D+Q^2]$.
     The enhanced classical Hamiltonian is given by
     \bn H(p,q)\hskip-0.8em &&=\half\langle p,q| DQ^{-2}D+Q^2|p,q\rangle\no \\ &&
      =\half\langle\beta| (D+pqQ)(qQ)^{-2}(D+pqQ)+(qQ)^2 |\beta\rangle \no \\ && =\half[p^2+q^2(1+{\cal{O}}(\hbar))+F(\hbar)/q^2]\,, \en
          where $F(\hbar)=\langle\beta|DQ^{-2}D|\beta\rangle>0$.
          The conventional classical (c) result is given by
          \bn H_c(p,q)=\lim_{\hbar\rightarrow 0} H(p,q)=\half[p^2+q^2] \,\,,\,\,\,q>0 \en
          as expected.
     

 \subsection{Schr\"odinger's representation and equation}
 The Schr\"odinger representation for $Q>0$ is $x>0$, where $x$ is simply a positive real number, and for $D$ is $-(i\hbar/2)[x(\partial/
 \partial x)+(\partial/\partial x)x]$ or $-i\hbar[x(\partial/\partial x) +1/2]$. Wave functions are   $\psi(x)$, which may be 
 normalized: $\tint_0^\infty |\psi(x)|^2\;dx=1$. For the problem at hand,
 Schr\"odinger's equation is given by
   \bn i\hbar\,\partial\psi(x,t)/\partial t\hskip-1em&&=\half\{-\hbar^2 [x(\partial/\partial x) +1/2] x^{-2}
   [x(\partial/\partial x) +1/2]\no \\
   &&\hskip6em+m_0^2\,x^2\}\;\psi(x,t) \;\\ 
           &&=\half\{-\hbar^2\partial^2/\partial x^2+(3/4)\hbar^2\,x^{-2} + m_0^2\,x^2\}\;\psi(x,t)\;. \no \en

We will continue to focus on Schr\"odinger's representation to present the
operators and relevant equations for the remaining examples.

 \section{A Field Theory Example}
 \subsection{Canonical quantization}
 As a more complex example we suggest the classical Hamiltonian, where $x\in\mathbb{R}$, given by
    \bn H(\pi,\phi) =\textstyle{\int}\{ \phi(x)\pi(x)^2+ (\nabla{\phi})(x)^2\}\;dx \,\;, \en
    subject to the restriction that $\phi(x)>0$. These fields satisfy the Poisson bracket
    $\{\phi(x),\pi(x')\}=\delta(x-x')$.
    Canonical quantization promotes $\pi(x)$ to 
    the operator $\hat{\pi}(x)$ and $\phi(x)$ to the operator $\hat{\phi}(x)>0$, which obey the
    commutator $[\hat{\phi}(x), \hat{\pi}(x')]=i\hbar\,\delta(x-x')\mathbb{I}$.
    
    These two field operators should be self-adjoint operators when smeared with test functions. However, the positivity of $\phi$ means that $\pi$ cannot be self adjoint. This situation complicates canonical quantization, and we do not discuss it further. We will find that affine quantization is more friendly!
    
    \subsection{Affine quantization}
    To proceed we introduce the classical affine field
     $\kappa(x)\equiv \pi(x)\phi(x)$, with $\phi(x)>0$,
      and observe that the principal Poisson bracket is given by
     \bn \{\phi(x),\kappa(x')\}=\delta(x-x')\,\phi(x)\;\;, \en
     with $\phi(x)>0$. These classical fields are promoted to operators such that the principal commutator is
       \bn [\hat{\phi}(x), \hat{\kappa}(x')]=i\hbar\,\delta(x-x')\,\hat{\phi}(x)\;,\en
       subject to the condition that $\hat{\phi}(x)>0$.

      \subsection{Affine coherent states}
      The affine coherent states for this model are given by
      \bn |\pi, \p\>=e^{(i/\hbar)\tint \pi(x)\hat{\phi}(x)\, dx}\,e^{-(i/\hbar)\tint\ln[\phi(x)]
      \hat{\kappa}(x)\,dx}\;|\nu\>\;.\en
    The fiducial vector is formally given by
       \bn \Pi_x\;[(\hat{\phi}(x)-1)+i\hat{\kappa}(x)/\nu\hbar]\;|\nu\>=0\;.\en
       This relation leads to $\<\nu|\hat{\p}(x)|\nu\>=1$ and $\<\nu|\hat{\kappa}(x)
       |\,\nu\>=0$.
       
       As before, and for a suitable factor $K$, the affine coherent states generate a resolution of unity such as
       \bn \tint|\pi,\p\>\<\pi,\p|\;\Pi_x\;d\pi(x)\,d\p(x)/2\pi\hbar K=\mathbb{I}\;,\en
 and for the present set of coherent states, the Fubini-Study metric 
 becomes
       \bn d\sigma(\pi,\p)^2\hskip-1.5em&&\equiv 2\hbar\tint[\,|\!|\,d|\pi(x),\p(x)\>|\!|^2-|\<\pi(x),\p(x)|\,d|\pi(x),\p(x)\>|^2]\;dx  \no \\
       &&=\tint[(\nu\hbar)^{-1}\p(x
       )^2\, d\pi(x)^2+(\nu \hbar)\p(x)^{-2}\,d\p(x)^2\,]\;dx\;.\en
       Evidently, this affine metric is an infinite set of separate constant negative curvature spaces, specifically $-2/\nu\hbar$, for every value of $x$. It is also noteworthy that the product of the coefficients of the two differential terms, i.e., coefficients of 
       $d\pi(x)^2$ and $d\p(x)^2$, is {\it unity} for all $x$. 
       
\subsection{Enhanced classical operators}
          Following the single degree of freedom model, it follows that
          \bn \<\pi,\p| \hat{\p}(x) |\pi,\p\>=\<\nu|\p(x)\,\hat{\p}(x)|\nu\>= \p(x)\;,\en and
      \bn \<\pi,\p| \hat{\kappa}(x) |\pi,\p\>\hskip-1em&&=\<0,\p| \pi(x)\,\hat{\p}(x))|0,\p\>\no \\ &&
      =\<\nu|\pi(x)\p(x)\hat{\p}(x)|\nu\>=\pi(x)\p(x)\;.\en
      
      The enhanced classical Hamiltonian for this model is given by
      \bn H(\pi,\p)=\<\pi,\p|\{\tint [\hat{\kappa}(x)\hat{\p}(x)^{-1}\hat
      {\kappa}(x)+ 
      (\nabla{\hat{\p}})(x)^2\,]\,dx\;\}|\pi,\p\>\;.\en
       This relation `reduces' so that
       \bn H(\pi,\p)\hskip-1em&&=\tint \{\<\nu|[\hat{\kappa}(x)+\pi(x)\p(x)\hat{\p}(x)]
       [\p(x)\hat{\p}(x)]^{-1}\no\\ \hskip4em&&\times
       [\hat{\kappa}(x)+\pi(x)\p(x)\hat{\p}(x)]+
       (\nabla
       [\p(x)\hat{\p}(x)])^2\}|\nu\>\} \;dx\;.\en
       Finally, we find that 
       \bn H(\pi,\p)=\tint[\p(x)\pi(x)^2+(\nabla{\p})(x)^2(1+{\cal{O}}(\hbar))+G(\hbar)/\p(x)]\;dx\;,\en
       where $G(\hbar)=\<\nu|\hat{\kappa}(x)\hat{\p}(x)^{-1}\hat{\kappa}
       (x)|\nu\> >0$ and $G(\hbar)\propto \hbar^2$.
     The truly classical expression becomes 
     \bn H_c(\pi,\p)=\lim_{\hbar\rightarrow0} H(\pi,\p)=\tint[\p(x)\pi(x)^2+(\nabla{\p})(x)^2]\;dx \;,\en
     with $\p(x)>0$, as expected.
     
  \subsection{Schr\"odinger's representation and equation}
      The Schr\"odinger representation for the principal operators is given by 
       $\hat{\phi}(x)= \phi(x)>0\;,$ and 
       \bn \hat{\kappa}(x)= -i(\hbar/2)[\phi(x)(\delta/\delta\phi(x))+(\delta/\delta\phi(x))\phi(x)]\;.\en
       The Hilbert state vectors are given as functionals such as $\Psi(\p)$, and are formally 
       normalized by $\tint|\Psi(\p)|^2\:{\cal{D}}\p =1$.
       Schr\"odinger's dynamical equation is now given by
       \bn i\hbar\,\d\Psi(\p,t)/\d t=\{\tint [\hat{\kappa}(x)\p(x)^{-1}\hat{\kappa}(x)+(\nabla{\p})(x)^2]
       \;dx\}\,\Psi(\p,t)\;.\en
       
       Regularization, by limiting the number
       of variables to a large but finite number of variables, enables such expressions to be 
       investigated more clearly.

 \section{Quantum Gravity}
 The present paper is intended to be an introductory story to a more complete and higher level version of the study of quantum gravity that is available in \cite{bqg}. That article offers more complex examples and the reader may wish to start with the more accessible story offered in the present article. In our present study we emphasize the importance of choosing an affine quantization rather than a conventual canonical
  quantization. In so doing, we will not offer here a complete story of quantizing gravity but rather focus on some basic issues; to see a more complete story, the article \cite{bqg} can be recommended.
 
 One of the most difficult problems in quantum gravity deals with the choice of classical variables picked to promote to quantum operators. To briefly see that problem up close we first choose the classical phase space version of the gravitational action functional described in \cite{adm}. In particular, we introduce
 \bn A=\tint_0^T\tint \{ -g_{ab}\dot{\pi}^{ab}-NH -N^aH_a\} \; d^3\!x\,dt\;. \en
 In this expression we have the symmetric-index metric field $g_{ab}( x,t)$, which is constrained 
 such that as a $3\times3$ matrix it is strictly positive as denoted by $\{g_{ab}(x,t)\}>0$, along with the symmetric-index
 momentum field $\pi^{ab}(x,t)$ (which has a time derivative here). We concentrate on a 3-dimensional space and indices
 $a,b,\cdots=1,2,3$ which are automatically summed in pairs. The positive metric tensor leads to the determinant $g(x,t)\equiv \det[g_{ab}(x,t)]$ which will also be positive, i.e., $g(x,t)>0$ for all $(x,t)$.
 The other terms are $H_a(x,t)=\pi^b_{a\,|b}(x,t)$, which is the diffeomorphism constraint and involves a covariant derivative of the mixed-index, momentum tensor given by
 $\pi^b_a(x,t)\equiv \pi^{bc}(x,t)\,g_{ac}(x,t)$. This mixed-index tensor will be a very important variable in our analysis, and we name it the momentric field which stems from the {\it momen}tum and me{\it tric} components. The Hamiltonian density (at fixed time) is given by
 \bn H(x)=g(x)^{-1/2}[\pi^a_b(x)\pi^b_a(x)-\half\pi^a_a(x)\pi^b_b(x)]+
     g(x)^{1/2}\,^{(3)}\!R(x)\;, \en
    where $ ^{(3)}\!R(x)$ is the spatial, three-dimensional scalar curvature.
    The remaining terms, $N^a(x,t)$ and $N(x,t)$, are Lagrange multipliers which enforce the set of four constraints, namely $H_a(x,t)=0$ and $H(x,t)=0$ for all $(x,t)$. 
    
    In a nutshell, this survey outlines the various quantities that determine the properties of classical gravity. 
    
    \subsection{Basic variables}
    The phase space variables (at a fixed time) are the metric field 
    $g_{ab}(x)$ and the momentum field $\pi^{cd}(x)$. These variables have a Poisson bracket given by $\{g_{ab}(x),\pi^{cd}(x')\}=\half\delta^3(x- x')[\delta^c_a\delta^d_b+\delta^d_a\delta^c_b]$.
    With the requirement that the proposed self-adjoint metric operators satisfy  $\{\hat{g}_{ab}(x)\}>0$, as well as $\hat{g}(x)\equiv\det[\hat{g}_{ab}(x)]>0$, it is impossible to ensure that any proposed momentum operator $\hat{\pi}^{cd}(x)$ could become self adjoint. This situation means that canonical quantization is difficult if not appropriate. To remedy this situation, different variables have been 
    chosen as new,
    basic canonical operators. That may resolve the mathematical issues, but the new variables, which may involve aspects of the metric and/or momentum, would most likely fail the implicit requirement of being `Cartesian coordinates'.\footnote{An aside: By analogy, the present situation would be like quantizing an harmonic oscillator, which normally maps the classical Hamiltonian $\half[p^2+q^2]$ to the quantum Hamiltonian $\half[P^2+Q^2]$. Instead, someone first changes the classical canonical  coordinates, e.g., $p=\bar{p}/\bar{q}^2$ and $q=\bar{q}^3/3$, and next, takes the classical Hamiltonian $\bar{H}(\bar{p},\bar{q})$
    expressed in the new variables, which are no longer `Cartesian', and then promotes it directly 
    to $\hat{\bar{H}}(\bar{P}, \bar{Q})$. The new operators would satisfy 
    $[\bar{Q},\bar{P}]=i\hbar\mathbb{I}$, and likely be self adjoint, but clearly the spectrum of the Hamiltonian operator would be different and thus the
    physics would be different.}
   
   Let us now entertain an affine point of view. The classical momentric 
   field, $\pi^a_b(x)\equiv \pi^{ac}(x)\,g_{bc}(x)$, and the metric field, $g_{ab}(x)$, become the new basic variables. These two variables have a joint set of Poisson brackets given by
   \bn &&\{\pi^a_b(x),\pi^c_d(x')\}=
   \half\,\delta^3(x-x')\s[\delta^a_d\s \pi^c_b(x)-\delta^c_b\s \pi^a_d(x)\s]\;,    \no \\
       &&\hskip-.20em\{g_{ab}(x), \s \pi^c_d(x')\}= \half\,\delta^3(x-x')\s [\delta^c_a g_{bd}(x)+\delta^c_b g_{ad}(x)\s] \;,      \\
       &&\hskip-.30em\{g_{ab}(x),\s g_{cd}(x')\}=0 \;. \no  \en
        Unlike  their canonical version, these Poisson brackets suggest that they are equally   valid if $g_{ab}(x)\ra -g_{ab}(x)$, and thus there could be separate realizations of each choice. 
        
Passing to operator commutators, promoted from the Poisson brackets, we are led to consider
 \bn   &&[\hp^a_b(x),\s \hp^c_d(x')]=i\s\half\,\hbar\,\delta^3(x-x')\s[\delta^a_d\s \hp^c_b(x)-\delta^c_b\s \hp^a_d(x)\s]\;,    \no \\
       &&\hskip-.10em[\hg_{ab}(x), \s \hp^c_d(x')]= i\s\half\,\hbar\,\delta^3(x-x')\s [\delta^c_a \hg_{bd}(x)+\delta^c_b \hg_{ad}(x)\s] \;, \\
       &&\hskip-.20em[\hg_{ab}(x),\s \hg_{cd}(x')] =0 \;. \no  \en
Indeed, there are two irreducible representations of the metric tensor operator: one where the matrix $\{\hg_{ab}(x)\}>0$, which we accept, and one where the matrix $\{\hg_{ab}(x)\}<0$, which we reject. After being smeared with suitable
   test functions, the result is that both the metric and the momentric tensors can be self-adjoint operators, and the metric operators will satisfy the required positivity requirements. It appears that affine  quantization has accomplished something that canonical quantization could not do!
   
Of course, critics may suggest that these operators may not have been promoted from 
favorable classical coordinates. In the next subsection we will show that
the basic affine gravity operators are indeed promoted from favored 
coordinates, just like the earlier examples in this paper.

    \subsection{Affine coherent states for gravity}
    We choose the basic affine operators to build our coherent states for gravity; specifcally,
   \bn |\pi,\eta\>=e^{(i/\hbar)\tint \pi^{ab}(x)\hat{g}_{ab}(x)\; dx} \; e^{-(i/\hbar)\tint\eta^a_b(x)\hat{\pi}^b_a(x)\;dx}\;|\alpha\>\;\;\;\;[=|\pi,g\>]. \en
   The fiducial vector $|\alpha\>$ has been chosen so that the matrix $\eta(x)\equiv\{\eta^a_b(x)\}$ enters the coherent states solely in the form given by
          \bn \langle\pi,\eta|\hat{g}_{ab}(x)|\pi,\eta\rangle =[e^{\eta(x)/2}]^c_a\,\langle\alpha|\hat{g}_{cd}(x)\,|\alpha\rangle\,[e^{\eta(x)/2}]^d_b\equiv g_{ab}(x)\;.\label{mm} \en
   A companion relation is given by
    \bn \langle\pi,\eta|\hat{\pi}^a_b(x)|\pi,\eta\rangle
=\pi^{ac}(x)\,g_{cb}(x)\equiv\pi^a_b(x)\;,\en
which involves the metric result from (\ref{mm}). These relations permit us to rename the coherent states from $|\pi,\eta\>$ to $|\pi,g\>$.

               As a consequence, the inner product of two gravity coherent states is given by 
   \bn \langle\pi'',g''|\pi',g'\rangle\hskip-1.3em&&=\exp\Big{\{}\textstyle{-2\int}b(x)\,d^3x \\
   &&\hskip-3em \times\ln\big\{
\frac{\det\{\frac{1}{2}[ {g''}^{ab}(x)+{g'}^{ab}(x)]+i\frac{1}{2\hbar}b(x)^{-1}[{\pi''}^{ab}(x)-{\pi'}^{ab}(x)]\}}{\det[{g''}^{ab}(x)]^{1/2}
\,\,\det[{g'}^{ab}(x)]^{1/2}}  \big\} \Big\}\;.\no \en
Here the scalar density function $b(x)>0$ ensures the covariance of this expression.

To test whether or not we have `favorable coordinates' we examine, with a suitable factor $J$, the Fubini-Study metric, given by
 \bn d\sigma(\pi,g)^2\hskip-1.4em&&\equiv J\hbar[\,\|\,d|\pi,g\>\|^2-|\<\pi,g|\;d|\pi,g\>|^2\,] \no \\
     &&=\tint \{(b(x)\hbar)^{-1} g_{ab}(x)\,g_{cd}(x)\,d\pi^{bc}(x)\,d\pi^{da}(x) \\
     &&\hskip2em +(b(x)\hbar) \,g^{ab}(x)\,g^{cd}(x)\,dg_{bc}(x)\,dg_{da}(x)\}\; d^3\!x \;.\no \en
    This metric, like the one in the previous section, represents a multiple family of constant negative curvature spaces. The product of coefficients of the differential terms is proportional to a constant rather like the previous affine metric stories. Based on the previous analysis we accept that
    the basic affine quantum variables have been promoted from basic affine classical variables.

   \subsection{Schr\"odinger's representation and equation}
   The Schr\"odinger representation is given by $\hat{g}_{ab}(x)=g_{ab}(x)$ with $\{g_{ab}(x)\}>0$, as well as
   $\hat{g}(x)=g(x)=\det[g_{ab}(x)]>0$,
  \bn \hat{\pi}^a_b(x)=-\half i \hbar[g_{bc}(x)\,(\delta/\delta g_{ac}(x))+(\delta/\delta g_{ac}(x))
  \,g_{bc}(x)]\;, \en
  and a state functional is given by $\Psi(\{g\})$, the argument referring to the whole metric tensor rather than just the determinant. A normalization is formally given by $\tint_{+} |\Psi(\{g\})|^2
  {\cal{D}} \{g\}=1$, where the subscript $_+$ limits the integration to positive metric tensors, 
  $\{g_{ab}(x)\}>0$.
  
  Schr\"odinger's equation is given by
  \bn i\hbar\,\d \Psi(\{g\},t)/\d t\hskip-1em&& =\{\tint[\, \hat{\pi}^a_b(x)\,g(x)^{-1/2}\,\hat{\pi}^b_a(x)
  -\half \hat{\pi}^a_a(x) \,g(x)^{-1/2}\,\hat{\pi}^b_b(x) \no \\
  &&\hskip4em +
  g(x)^{1/2}\,^{(3)}\!R(x)\,]\;d^3\!x\,\}\;\Psi(\{g\},t) \label{seq}
    \;. \en
    We emphasize that solutions of this equation involve smooth metrics $g_{ab}(x)$ for all elemants
    at each value 
    of $x$; this ensures the finiteness of the metric derivatives at every point of space. This
    result is at odds with certain different formulations of quantum gravity.
    
    It is noteworthy that $\hat{\pi}^a_b(x)\,g(x)^{-1/2}=0$, which implies that the Hamiltonian density in (\ref{seq}) is given by
    \bn \mfH'(\hat{\pi}^a_b(x), g_{ab}(x))&&  \no  \\
    &&\hskip-5em =g(x)^{-1/2}[\hat{\pi}^a_b(x)\hat{\pi}^b_a(x)-\half
    \hat{\pi}^a_a(x)\hat{\pi}^b_b(x)] + g(x)^{1/2}\,^{(3)}\!R(x)\;,  \en
    and, when viewed as a constraint, leads to physical Hilbert space vectors $\Phi(\{g\})$ that satisfy
      \bn 0=\{[\hat{\pi}^a_b(x)\hat{\pi}^b_a(x)-\half
    \hat{\pi}^a_a(x)\hat{\pi}^b_b(x)]+g(x)\,^{(3)}\!R(x)\}\,\Phi(\{g\})\;.\label{uuu} \en
    An easy way to find solutions to this equation is to choose a smooth physical Hilbert state $\Phi(\{g\})$ function, and then use it to select $^{(3)}\!R(x)$. Moreover, if $\Phi(\{g\})$ is a solution so is $\Phi'(\{g\})=\Phi(\{g\})\Pi_x|g(x)|^{-1/2}$, modulo proper normalization.
    
    Of course, these expressions are made more manageable with suitable  regularization such as used in \cite{bqg}.

    \section{Summary, and Outlook}
  The equations above are fundamental to quantum gravity. Their validity for a large spatial realm implies
  the validity of every arbitrarily small region of space, as well as the validity of the quantum Hamiltonian density $\mfH(x,t)$. In a certain sense, the difficulty of defining the quantum Hamiltonian density function is the most difficult hurdle to overcome in any quantum gravity study. If a suitable examination of  
  (\ref{uuu}) reveals further positive results, then the doors are open to complete the study of quantum gravity. The ingredients of a successful quantization include enforcing the classical constraints,
  with proper treatment of second-class constraints if there are any,
  reducing the kinematical Hilbert space to the all-important physical Hilbert space, and calculating some issues of importance. 
  
  To go further there are a number of analytical suggestions available in \cite{bqg}. No doubt there are numerous other procedures in the literature that can flesh out one or another approach to advance the story of quantizing gravity.

\section*{Acknowledgements} Thanks to Ewa Czuchry for suggesting several clarifications of the
presentation.

\end{document}